\documentclass[twocolumn,twoside]{article}
\usepackage{epsfig}
\usepackage{graphicx}

\newcommand{\hb}{\\ \hspace*{2ex}}

\begin{document}
\title{HIGHLY RELATIVISTIC CIRCULAR ORBITS OF SPINNING PARTICLE IN THE  KERR FIELD }
\author{Roman Plyatsko and Mykola Fenyk\\[2mm]
 Pidstryhach Institute for Applied  Problems in Mechanics and Mathematics \\ of Ukrainian National Academy of Sciences,\hb
 Lviv, Ukraine
}
\date{}
\maketitle
ABSTRACT.
The Mathisson-Papapetrou equations in Kerr's background
 are considered. The region of existence of highly
relativistic planar circular orbits of a spinning particle in this
background and dependence of the particle's Lorentz $\gamma$-factor on its spin and radial coordinate are investigated. It is shown that in
contrast to the highly relativistic circular orbits of a spinless
particle the corresponding orbits of a spinning particle are allowed in much
wider space region. Some of these orbits show the significant attractive action of the spin-gravity coupling on a particle and others are caused by the significant repulsive action. Numerical estimates for electrons, protons and neutrinos in the gravitational field of black holes are presented. \\[1mm]


\section{ Introduction}

Last year, 2012, was the 75th anniversary of the important equations in gravitational physics. There are different forms of the name for these equations in the literature: {\it Mathisson-Papapetrou} equations, {\it Papapetrou}, {\it Mathisson-Papapetrou-Dixon}, {\it Papapetrou-Dixon}, and rarely {\it Mathisson} equations. In any case,  these equations first appeared in the paper by Mathisson \cite{Mathis1}. The second paper, where those equations were derived by the alternative method, was published by Papapetrou \cite{Papa}. Another approach to these equations was developed  W.G. Dixon \cite{Dixon1, Dixon2, Dixon3, Dixon4}, and  later many authors elaborated on this subject as well \cite{Tulcz, Taub, Bartrum, Kunzl, Omote, Hojman, Balach, Fuchs, Natario, Barausse, Stein1, Stein2, Gralla}. Below we shall use the term {\it Mathisson-Papapetrou} (MP) equations.

The MP equations describe motions of a classical (nonquantum) spinning particle in the pole-dipole approximation in a gravitational field according to general relativity. In the aforementioned papers of Dixon and others, the generalization of these equations for the higher multipoles were developed, and this generalization are often dubbed the {\it Mathisson-Papapetrou-Dixon} (or simply {\it Dixon}) equations.

Note that the equations for a quantum particle with spin 1/2 in the gravitational field are eight years "older" than the MP equations. Indeed, in \cite{Fock1, Fock2, Weyl} the usual Dirac equation was generalized for curved spacetime. Much later the connection between this equation and the MP equations was investigated in some papers \cite{Wong, Kann, Caten, Aud, Gorb, Barut, Cian1, Cian2, Obukh}  and it was shown that in a certain sense the MP equations follow from the general relativistic Dirac equation as a classical approximation.

The MP equations are considered with some supplementary condition in order to choose an appropriate representative world line which can be applied to describe the spinning particle trajectory. Different conditions are used in the literature, and the Mathisson-Pirani \cite{Mathis1}, Tulzcyjew-Dixon \cite{Tulcz, Dixon1}, and Corinaldesi-Papapetrou \cite{CorPapa} conditions are best known. The vast list of  publications where different variants of the supplementary conditions were taken into account is presented in \cite{Semer, Kyr}. The first effects of the spin-gravity interaction following from the MP equations  in Schwarzschild's background were considered in \cite{CorPapa} at the Corinaldesi-Papapetrou condition. According to \cite{CorPapa} the influence of spin on the particle's trajectory is too small for practical registration. A similar conclusion was stressed in the known book \cite{Misner} where the MP equations under the Mathisson-Pirani conditions were discussed. Nevertheless,  another supposition can be found in \cite{Rasb}:  "{\it The simple act of endowing a black hole with angular momentum has led to an unexpected richness of possible physical phenomena. It seems appropriate to ask whether endowing the test body with intrinsic spin might not also lead to surprises}". Paper \cite{Rasb} together with \cite{Wald}, where the spin-spin and spin-orbit gravitational interactions were considered, gave the impulse for realizing  the program of more detailed investigations of physical effects following from the MP equations without {\it a priori} restrictions on the influence of the particle's spin on its trajectory.  One of the first results of this program realization was presented in \cite{DAN}, where the specific nonequatorial highly relativistic circular orbits of a spinning particle in the Kerr background were studied.

Different highly relativistic circular orbits are considered in investigations of possible synchrotron radiation, both electromagnetic and gravitational, of protons and electrons in the gravitational field of a black hole \cite{Bre, Chrz, Sai, Suz, Moh, Pl1}. In the theoretical plane, it is known that the highly relativistic circular orbits of a spinless particle are  important in the classification of all possible geodesic orbits in a Kerr spacetime. Similarly, the highly relativistic circular orbits of a spinning particle are important in the classification of all possible nongeodesic orbits in this spacetime as well.

The circular orbits of a spinning particle according to the MP equations in the Schwarzschild, Kerr and other backgrounds were considered in many papers \cite{Tod, Hoj2, Abr, Svir, Suz2, Bini1, Far, Bini2, Bini3, Bini4, Bini5, Bini6, Bini7, Moh2}, and very highly relativistic orbits  were studied in \cite{Pl2, Pl3}. The purpose of this paper is to present the results of investigations of highly relativistic equatorial circular orbits in Kerr's background which follow from the MP equations. Note that in \cite{Pl2} the corresponding orbits were considered only near $r_{ph}^{(-)}$, in some narrow space region (here $r_{ph}^{(-)}$ is the Boyer-Lindquist radial coordinate of the counter-rotating circular photon orbits). In \cite{Pl3} we dealt with the Schwarzschild background only. That is, the present paper can be considered as a development and generalization of \cite{Pl2, Pl3}. The main features of the spin-gravity interaction that are revealed on the circular orbits will be a good reference to further investigations of most general motions of a highly relativistic spinning particle in the Kerr background.

The title of Mathisson's paper \cite{Mathis1} in English is {\it New Mechanics of Material Systems}. To understand the sense of these words it is necessary to know the main features of a single spinning particle's motions in the gravitational field according to the MP equations, first of all at its high velocity.

In Sec. II,  basic information about the MP equations and their supplementary conditions is presented. The concrete form of the algebraic equations which follow from the exact MP equations for the circular motions of a spinning particle in the Kerr background is written in Sec. III. The results of solutions of those equations for different cases of the spin orientation and the direction of the particle's orbital motion on the circular orbits are described in Secs. IV and V. We conclude in Sec. VI.

\section{Mathisson-Papapetrou equations and supplementary conditions}

 The traditional form of  MP equations is \cite{Mathis1, Papa}
\begin{equation}\label{1}
\frac D {ds} \left(mu^\lambda + u_\mu\frac {DS^{\lambda\mu}}
{ds}\right)= -\frac {1} {2} u^\pi S^{\rho\sigma}
R^{\lambda}_{~\pi\rho\sigma},
\end{equation}
\begin{equation}\label{2}
\frac {DS^{\mu\nu}} {ds} + u^\mu u_\sigma \frac {DS^{\nu\sigma}}
{ds} - u^\nu u_\sigma \frac {DS^{\mu\sigma}} {ds} = 0,
\end{equation}
where $u^\lambda\equiv dx^\lambda/ds$ is the particle's 4-velocity,
$S^{\mu\nu}$ is the tensor of spin, $m$ and $D/ds$ are,
respectively, the mass and the covariant derivative with respect to
the particle's proper time $s$ and $R^{\lambda}_{~\pi\rho\sigma}$ is
the Riemann curvature tensor (units $c=G=1$ are used). Here, and in
the following, Latin indices run 1, 2, 3 and Greek indices 1, 2, 3,
4; the signature of the metric (--,--,--,+) is chosen.

In order to describe motions of a concrete point which represents the motions of the spinning particle as a whole, it is necessary to supplement
Eqs.  (\ref{1}) and  (\ref{2}) with an additional relationship. The natural point is the particle's center of mass, and for its determination most often the Mathisson-Pirani or Tulczyjew-Dixon conditions are used. The Mathisson-Pirani condition is the relationship between $S^{\lambda\nu} $ and $u_\nu$
\cite{Mathis1, Pirani}
\begin{equation}\label{3}
S^{\lambda\nu} u_\nu = 0.
\end{equation}
In the Tulczyjew-Dixon condition instead of  $u_\nu$ the particle's 4-momentum  $P_\nu$ is written \cite{Tulcz, Dixon1}:
\begin{equation}\label{4}
S^{\lambda\nu} P_\nu = 0,
\end{equation}
where
\begin{equation}\label{5}
P^\nu = mu^\nu + u_\lambda\frac {DS^{\nu\lambda}}{ds}.
\end{equation}

In general, $P_\nu$ is not parallel to $u_\nu$ and conditions  (\ref{3}) and  (\ref{4}) are different. Correspondingly, in spite of the fact that in some cases the solutions of the MP equations under conditions (\ref{3}) and  (\ref{4}) are close or even coincide, in general they are different.

Below in our paper we use the the Mathisson-Pirani condition. This choice is determined by its physical meaning. First, as it is stressed in \cite{Taub}, the MP condition arises in a natural fashion in the course of the derivation. Second, we take into account the well-argued conclusion from  \cite{Costa1} concerning the clear and correct physical meaning of this condition, as well as the statement that {\it "Many applications in this work are examples where the  Mathisson-Pirani condition is the best choice ... "} \cite{Costa2}. One can find more detailed discussion on this condition in the Introduction of paper  \cite{Pl3}.

Equations  (\ref{1}) and  (\ref{2}) have the constant of motion
\begin{equation}\label{6} S_0^2=\frac12
S_{\mu\nu}S^{\mu\nu},
\end{equation}
where $|S_0|$ is the absolute value of spin. While dealing with the MP equations in the Schwarzschild and Kerr metrics with the aim to study possible physical effects caused by spin-gravity interaction, it is necessary to take into account  the condition for a spinning test particle \cite{Wald}
\begin{equation}\label{7}
\frac{|S_0|}{mr}\equiv\varepsilon\ll 1,
\end{equation}
 where $r$ is  the radial coordinate. For a macroscopic spinning test particle relationship (\ref{7}) is a direct consequence  of the fact that for this particle $|S_0|$ is of the order $m v_{rot} r_{part}$, where $v_{rot}$ is the rotation velocity of a point at the particle's surface and $r_{part}$ is the radius of the particle, with the clear conditions $v_{rot}<1$ and $r_{part}/r\ll 1$.

\section{ Equations for circular orbits in Kerr's background}

For the general type of a spinning particle motions in the Kerr background, the MP equations are a complicated system of  differential equations
\cite{Pl4}. In the partial case of the circular orbits from those equations,  some algebraic relationships follow, which we consider in this section.

Let us use the Kerr metric in the
Boyer-Lindquist coordinates  $x^1=r, \quad x^2=\theta, \quad
x^3=\varphi, \quad x^4=t$ with the nonzero components of the metric tensor $g_{\mu\nu}$ as
\[
g_{11}=-\frac{\rho^2}{\Delta}, \quad g_{22}=-\rho^2,
\]
\[
g_{33}=-\left(r^2+a^2+\frac{2Mra^2}{\rho^2}
\sin^2\theta\right)\sin^2\theta,
\]
\begin{equation}\label{8}
 g_{34}=\frac{2Mra}{\rho^2}\sin^2\theta, \quad
g_{44}=1-\frac{2Mr}{\rho^2},
\end{equation}
where
\[
 \rho^2=r^2+a^2\cos^2\theta, \quad \Delta=r^2-2Mr+a^2, \quad
0\le\theta\le\pi.
\]
(In the following, we shall put $a\ge 0$, without any loss in
generality.)
Note  that the corresponding expressions for the Christoffel symbols and the Riemann tensor components in metric  (\ref{8}) can be found, for example, in \cite{Semer}.

Together with the tensor of spin $S^{\mu\nu}$, for different purposes the 4-vector of spin $s_\lambda$ and the 3-vector $S_i$ are used as well, where by definition
\begin{equation}\label{9}
s_\lambda=\frac12 \sqrt{-g}\varepsilon_{\lambda\mu\nu\sigma}u^\mu
S^{\nu\sigma}
\end{equation}
and
\begin{equation}\label{10}
S_i=\frac{1}{2u_4} \sqrt{-g}\varepsilon_{ikl}S^{kl},
\end{equation}
here $g$ is the determinant of the metric tensor,
$\varepsilon_{\lambda\mu\nu\sigma}$ and  $\varepsilon_{ikl}$ are the
spacetime and spatial Levi-Civita  symbols, respectively. There is
the relationship between $S_i$ and $s_\lambda$:
\begin{equation}\label{11}
S_i=-s_i+\frac{u_i}{u_4}s_4.
\end{equation}

Let us consider the possible equatorial circular orbits of a spinning particle in the equatorial plane $\theta=\pi/2$ of the Kerr source with
 \begin{equation}\label{12}
 u^1=0, u^2=0, u^3=const\ne 0, u^4=const\ne 0,
\end{equation}
when spin is orthogonal to this plane, with
\begin{equation}\label{13}
S_1\equiv S_r=0, \quad S_2\equiv S_{\theta}\ne 0, \quad S_3\equiv S_{\varphi}=0.
\end{equation}
Then it is not difficult to show that by (\ref{8}), (\ref{12}) and (\ref{13}) at condition (\ref{3}) from Eqs. (\ref{2})  follows the single nontrivial relationship
\begin{equation}\label{14}
S_\theta=u_4 S_0.
\end{equation}
As a result, among the four Eqs. (\ref{1}) only the first of them is nontrivial if the parametrization condition $u_\mu u^\mu=1 $ is taken into account.
(Note that in the general case of any metric, the MP equations have the constant of motion $u_\mu u^\mu=const $.] We write this equation using, as in \cite{Pl2, Pl4}, the dimensionless quantities $y_i$ connected with the particle's coordinates and 4-velocity by definition
$$
y_1=\frac{r}{M},\quad y_2=\theta,\quad y_3=\varphi, \quad
y_4=\frac{t}{M},
$$
\begin{equation}\label{15}
y_5=u^1,\quad y_6=Mu^2,\quad y_7=Mu^3,\quad y_8=u^4.
\end{equation}
Then the mentioned equation from set  (\ref{1}) can be written as
$$
(\alpha^2 - y_1^3)y_7^2 - 2\alpha y_7 y_8 + y_8^2 -
3\alpha \varepsilon_0(y_1^2+\alpha^2)y_7^2 y_1^{-2}
$$
$$
+3\varepsilon_0(y_1^2+2\alpha^2)y_7 y_8 y_1^{-2} -
3\alpha \varepsilon_0 y_8^2 y_1^{-2}
$$
$$
+\alpha \varepsilon_0(3y_1^2+\alpha^2)(y_1^3-\alpha^2)y_7^4 y_1^{-3}-
\alpha \varepsilon_0\left(1-\frac{2}{y_1}\right)y_8^4 y_1^{-3}
$$
$$
+\varepsilon_0(y_1^6-3y_1^5 - 3\alpha^2 y_1^3 + 9\alpha^2 y_1^2 + 4\alpha^4)y_7^3 y_8 y_1^{-3}
$$
$$
+\alpha \varepsilon_0(3y_1^3-11y_1^2 - 6\alpha^2 + 2\alpha^2 y_1^{-1})y_7^2 y_8^2  y_1^{-3}
$$
\begin{equation}\label{16}
+\varepsilon_0(4\alpha^2-4\alpha^2 y_1^{-1} - y_1^3+3y_1^2)y_7 y_8^3 y_1^{-3}=0,
\end{equation}
where
\begin{equation}\label{17}
\varepsilon_0\equiv \frac{S_0}{mM}, \quad \alpha\equiv \frac{a}{M},
\end{equation}
and $a$ is the Kerr parameter from  (\ref{8}). In contrast to the value $\varepsilon$ from (\ref{7}), which depends on the radial coordinate, the value $\varepsilon_0$ from  (\ref{17}) is constant, and thus it is more convenient in our calculations. Below we put $|\varepsilon_0|\ll 1$, then condition  (\ref{7}) is satisfied for any $r$ of the order $m$ and larger.
It is easy to check that for $\varepsilon_0=0$ Eq.  (\ref{16}) corresponds to the equation which follows from the geodesic equations for the circular orbits of a spinless particle in Kerr's background in the Boyer-Lindquist coordinates.

Note that among eight values of $y_i$ from  (\ref{15}) only three of them, namely, $y_1$, $y_7$ and $y_8$, are present in Eq.  (\ref{16}), because per (\ref{12}) for the circular motions we have $y_5=0$ and $y_6=0$. In addition to Eq.  (\ref{16}), the values  $y_1$, $y_7$, and $y_8$ are connected by the relationship
$$
- \left(y_1^2 + \alpha^2 + \frac{2\alpha^2}{y_1}\right)y_7^2 + \frac{4 \alpha y_7 y_8}{y_1}
$$
\begin{equation}\label{18}
+\left(1-\frac{2}{y_1}\right)y_8^2=1
\end{equation}
which follows directly from the condition  $u_\mu u^\mu=1 $ by using the notation  (\ref{15}).

Therefore, for any fixed value of the radial coordinate, i.e., $y_1$, we have the two algebraic equations  (\ref{16}) and  (\ref{18}) which let us find the values of $y_7$ and $y_8$, in particular the values of the orbital particle's velocity, which are necessary for the motions on the possible circular orbits. We shall solve Eqs. (\ref{16}) and  (\ref{18}) using computer calculations. The corresponding results are presented in the next sections.

It is known that the geodesic equations in Kerr's background admit the highly relativistic circular orbits of a particle with the nonzero mass only in the small neighborhood of the values  $r_{ph}^{(+)}$ and  $r_{ph}^{(-)}$
that are the radial coordinates of the co-rotating and counter-rotating circular photon orbits, and these values are determined by the algebraic equation
\begin{equation}\label{19}
r(r-3M)^2 - 4Ma^2=0.
\end{equation}

Similarly to \cite{Pl2, Pl3}, the figures below show the dependence of the relativistic Lorentz $\gamma$-factor of a moving particle as calculated by an observer which is at rest relative to the Kerr mass. In terms of (\ref{15}) and (\ref{17}) the expression for this $\gamma$-factor for any circular motion in the equatorial plane is
\begin{equation}\label{20}
\gamma=\left(1-\frac{2}{y_1}\right)^{1/2}\left(y_8+\frac{2\alpha y_7}{y_1-2}\right).
\end{equation}
That is, Eq.  (\ref{20}) follows from the general expression for the 3-velocity
components $v^i$ of a particle moving relative to an observer which is at rest relative to Kerr's mass:
 \begin{equation}\label{20a}
v^i=\frac{dx^i}{\sqrt{g_{44}}}\left(dt+\frac{g_{4i}}{g_{44}}dx^i\right)^{-1}.
\end{equation}
Per  (\ref{20a}) it is easy to calculate the absolute value of this velocity $|v|$ using the expression
 \begin{equation}\label{20b}
\quad |v|^2=v_iv^i=\gamma_{ij}v^iv^j,
\end{equation}
where $\gamma_{ij}$ is the 3-space metric tensor which is connected with $g_{\mu\nu}$ by the known relationship
$$
\gamma_{ij}=g_{ij}+\frac{g_{4i}g_{4j}}{g_{44}}.
$$
In the case of the circular motions in the Kerr background, when $v^1=0$
 and $v^2=0$, according to  (\ref{20a}) we have
\begin{equation}\label{20c}
\quad
v^3=\frac{dx^3}{\sqrt{g_{44}}}\left(dt+\frac{g_{4i}}{g_{44}}dx^i\right)^{-1}=
\frac{u^3}{\sqrt{g_{44}}}\left(u^4+\frac{g_{43}}{g_{44}}u^3\right)^{-1}.
\end{equation}
Then Eq. (\ref{20}) can be obtained directly from the expression $\gamma=1/\sqrt{1-|v|^2}$, where relationships (\ref{20b}), (\ref{20c}),
and the explicit expressions for $g_{\mu\nu}$ from (\ref{8}) are taken into account.

\section{ Circular orbits with $a>0$, $S_\theta>0$ }

For an explanation of  all the circular motions of a spinning particle in the Kerr background, which are presented below in Figs. 1--16, it is necessary to compare these motions with some corresponding circular orbits of a spinning particle in Schwarzschild's background, which are illustrated in Figs. 1--5 of paper \cite{Pl3}. In addition, we take into account some features of the circular motions of a spinless particle in Kerr's background which are described by the geodesic equations. In particular, here we keep the terminology which is used traditionally for the two cases of the circular orbits: when the signs of $a$ and $d\varphi/ds$ are the same ({\it co-rotation}) and when these signs are different ({\it counter-rotation}). However, for the circular orbits of a spinning particle these cases include the two subcases, with the particle's spin "up" and "down". Correspondingly, in our consideration below it is convenient to present the figures and results as grouped by the two different orientations of the particle's spin. Namely, in this section we consider the case with   $S_\theta>0$, whereas in the next section we put $S_\theta<0$ (note that according to (\ref{14}) and (\ref{17}) the sign of $S_\theta$ coincides with the sign of $\varepsilon_0$).
Because it is pointed out above that we put $a\ge 0$, in these sections the cases of the circular orbits with $d\varphi/ds>0$ and $d\varphi/ds<0$ are
dubbed co-rotation and counter-rotation, respectively.

Before beginning the analysis of the graphs for Kerr's background which are presented in Figs. 1--16, let us recall the important features of the circular spinning particle orbits which were revealed in \cite{Pl3} for Schwarzschild's
background.  If $S_\theta>0$ and $d\varphi/ds>0$, at the large values of the relativistic Lorentz $\gamma$-factor the spin-gravity coupling causes an attractive action, in addition to the usual ("geodesic") attraction. Therefore, according to Figs. 1 and 2 from \cite{Pl3}, in the space region where the MP equations admit the circular orbits, the situation is possible when on the circular orbits with the same $r$ the spinning particle can move at the two different velocities: the first of them is close to the velocity of a spinless particle on the circular orbit with the same $r$, i.e. at this velocity the role of the spin-gravity coupling is small; the second value of the spinning particle velocity is much larger than the first, and in this case the role of the spin-gravity coupling is  important. We stress that if a spinless particle starts in the tangential direction with this large velocity at the same initial $r$, it begins a quick motion away from the Schwarzschild mass, i.e., at this velocity the usual gravitation cannot hold this spinless particle on the circular orbits.However, the spin-gravity coupling, which in the certain sense is proportional to $\gamma^2$ \cite{Pl5}, is able to perform this action.

If in Schwarzschild's background $S_\theta>0$ and $d\varphi/ds>0$, at the large values of the relativistic Lorentz $\gamma$-factor the spin-gravity coupling causes a significant repulsion action, and according to Figs. 4 and 5 from  \cite{Pl3}, due to this action the highly relativistic circular orbits of a spinning particle exist even in the space region where the circular orbits of a spinless particle are absent.

Now we can compare the new results on the highly relativistic circular orbits of a spinning particle in Kerr's background with the corresponding results for the Schwarzschild background.

\subsection{Case with $d\varphi/ds>0$ (co-rotation)}

In consideration of the possible highly relativistic circular orbits of a spinning particle, which follow from Eqs. (\ref{16})  and (\ref{18}), we begin from the co-rotation orbits with $S_\theta>0$, when for the chosen $a>0$ according to notation (\ref{15}) we put $y_7>0$. All the cases of the numerical calculations which are presented in Figs. 1--9 correspond to the value $\varepsilon_0=0.01$.

Figures 1--5 illustrate both the space domain of existence of the circular orbits and the dependence of the $\gamma$-factor on the radial coordinate for these orbits at different values of the Kerr parameter $a$. Figure 1 describes the case $a=0.0145M$, when $r_{ph}^{(+)}\approx 2.983M$ and $r_{ph}^{(-)}\approx 3.017M$. Note that for $r\leq r_{ph}^{(+)}$ there is no circular orbit of the spinning particle, i.e., this situation is the same as for the spinless particle. In the narrow space region between $r=r_{ph}^{(+)}$ and $r\approx 3.006M$ there are  highly relativistic circular orbits with a much higher Lorentz factor than  necessary for the spinless particle (for comparison, the dotted line in Fig. 1 shows the curve for the geodesic motion). We stress that the existence of those orbits is caused by the interaction of the particle's spin with the angular momentum of Kerr's source: in the Schwarzschild background the corresponding orbits are absent, according to Fig. 1 from  \cite{Pl3}.  Note that per the solutions of Eqs.  (\ref{16}) and (\ref{18}), the necessary value $\gamma$ tends to $\infty$ if $a$ tends to 0. In the wide space region for $r$ larger than $r\approx 3.006M$  Eqs.  (\ref{16}) and (\ref{18}) admit the two circular orbits for any fixed value $r$ and in this sense Fig. 1 below is similar to Fig. 1 from \cite{Pl3} for the Schwarzschild background.

For the values larger than $a\approx 0.0147M$ the curves $\gamma (r)$ significantly differ from the curves in Fig. 1. For example, as shows Fig. 2 for  $a=0.015$, the right branch of the curve in Fig. 1 is changed by the line which in the narrow space region goes to the corresponding geodesic line. According to Fig. 3,  for  larger $a$ the left-hand curve practically coincides with the geodesic line for all  values of $r>r_{ph}^{(+)}$, where
$r=r_{ph}^{(+)}$ is an asymptote. The right-hand curves in Figs. 2 and 3 are decreasing from the corresponding asymptotes in some space regions. However, for $r$ greater than $r\approx 4M$ these curves begin to grow, as  shown in Figs. 4 and 5 for different values $a$. Per Fig. 5 all the curves for $r$ greater than  $r\approx 8M$ tend to the curve for Schwarzschild's case which is presented in Fig. 2 from \cite{Pl3}. As  pointed out in \cite{Pl3}, in this case the curve $\gamma (r)$ for the greater $r$ is growing as $\sqrt{r}$.

We stress that all the curves $\gamma (r)$ for the spinning particle lay above the corresponding geodesic lines. This situation is similar to Schwarzschild's co-rotation case considered in \cite{Pl3}, where the spin-gravity interaction causes an additional attractive action as compare to the usual geodesic attraction. We also note that in the space domains where according to Figs. 1--5 there are two possibilities for a spinning particle's circular motions with the same $r$, the situation is similar to that described above for Schwarzschild's background: in the first case, the necessary velocity is close to the velocity of a spinless particle on the circular orbit with the same $r$, and here the spin-gravity effect is weak, whereas in the second possibility, with a much larger velocity, the influence of the spin-gravity coupling is  important.

\begin{figure}
[h] \centering
\includegraphics[width=5cm]{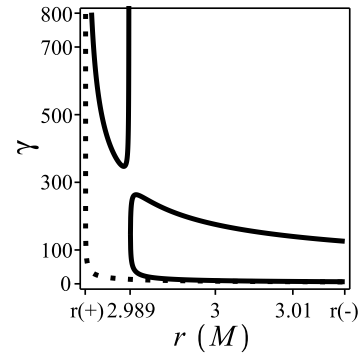}
\caption{\label{1} Dependence of the Lorentz factor on $r$ for the highly relativistic circular orbits with $d\varphi/ds>0$ of the spinning particle in Kerr's background at $a=0.0145M$, $\varepsilon_0=0.01$ (solid lines). The dotted line corresponds to the geodesic circular orbits. Here and in other figures below we use the notation $r(-)$ for  $r_{ph}^{(-)}$ and   $r(+)$ for  $r_{ph}^{(+)}$. }
\end{figure}

\begin{figure}
[h] \centering
\includegraphics[width=5cm]{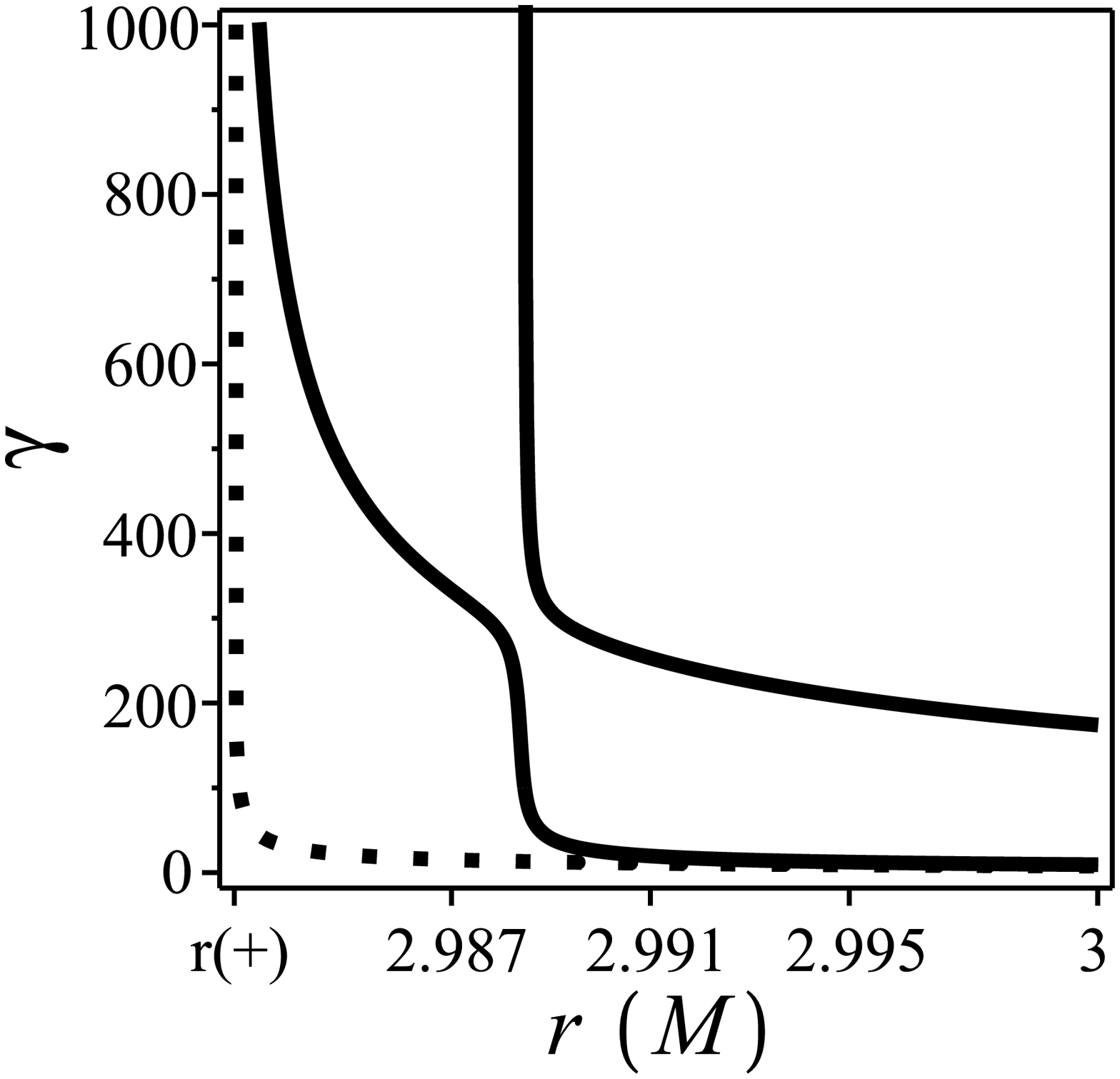}
\caption{\label{2} Dependence of the Lorentz factor on $r$ for the highly relativistic circular orbits with $d\varphi/ds>0$ of the spinning particle in Kerr's background at $a=0.015M$, $\varepsilon_0=0.01$ (solid lines). The dotted line corresponds to the geodesic circular orbits. }
\end{figure}

\begin{figure}
[h] \centering
\includegraphics[width=5cm]{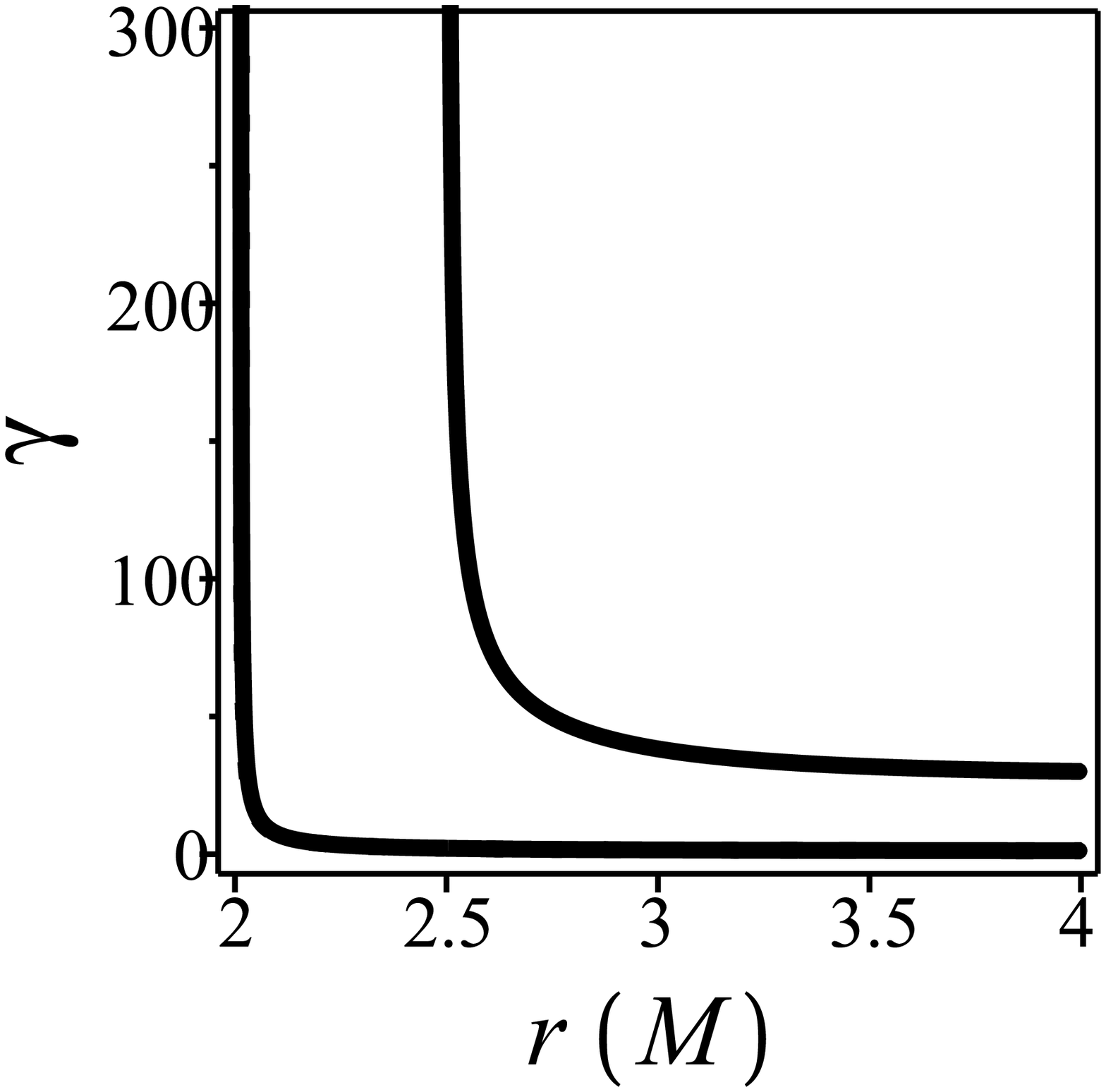}
\caption{\label{3} Lorentz factor vs $r$ for the highly relativistic circular orbits with $d\varphi/ds>0$ of the spinning particle in Kerr's background at $a=0.7M$, $\varepsilon_0=0.01$. }
\end{figure}

\begin{figure}
[h] \centering
\includegraphics[width=5cm]{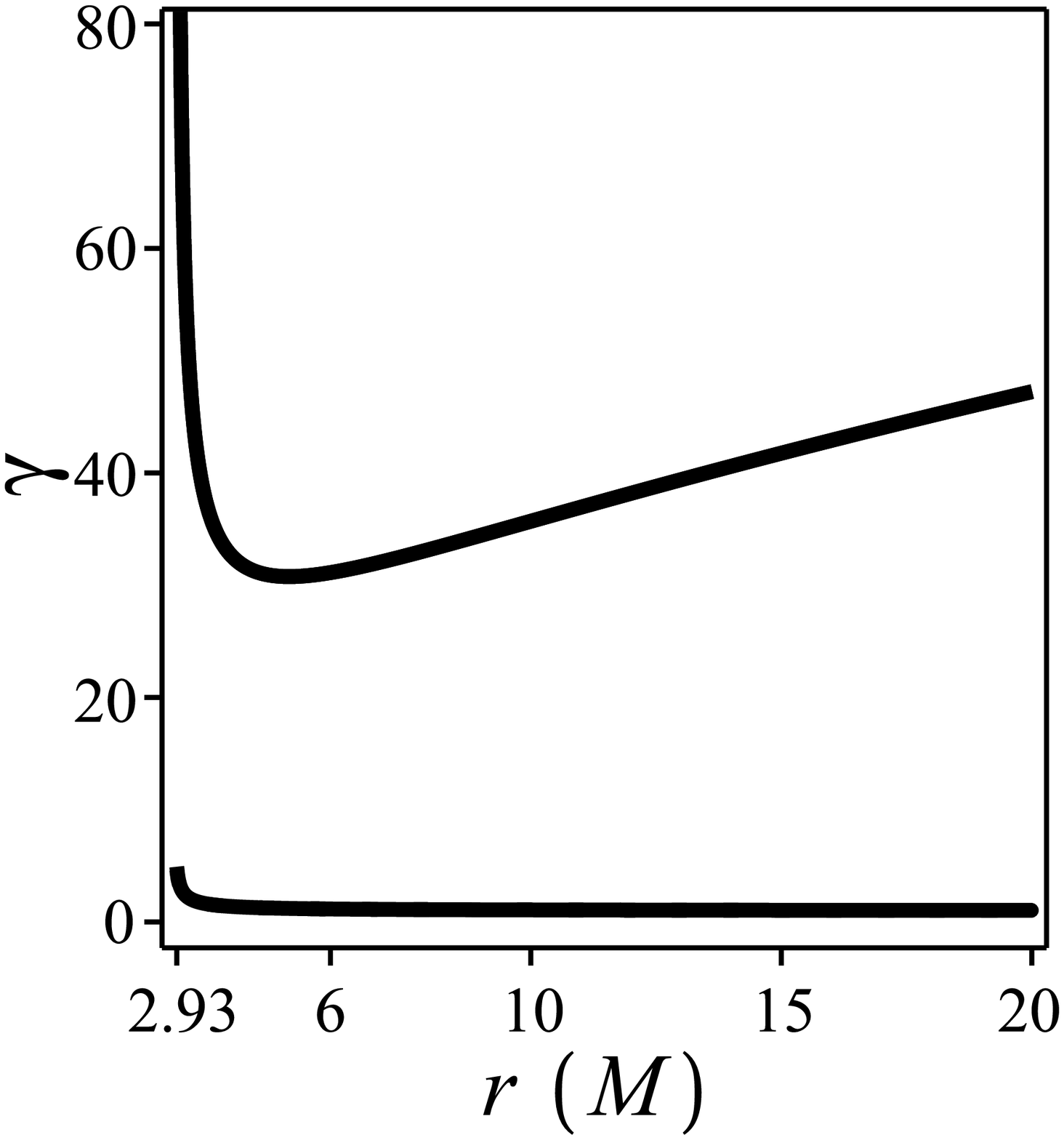}
\caption{\label{4} Lorentz factor vs $r$ for the highly relativistic circular orbits with $d\varphi/ds>0$ of the spinning particle in Kerr's background at $a=0.1M$, $\varepsilon_0=0.01$. }
\end{figure}

\begin{figure}
[h] \centering
\includegraphics[width=5cm]{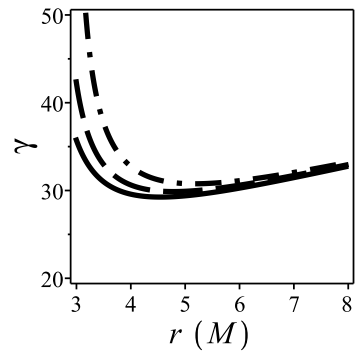}
\caption{\label{5} Lorentz factor vs $r$ for $\varepsilon_0=0.01$, $d\varphi/ds>0$ at $a=0.1M$ (dash-dotted line), $a=0.5M$ (dashed line), and $a=M$ (solid line). }
\end{figure}

\subsection{Case with $d\varphi/ds<0$ (counter-rotation)}

Figures 6--9 illustrate the circular orbits of the spinning particle which exist both for $r>r_{ph}^{(-)}$ and  $r\leq r_{ph}^{(-)}$, in contrast to the geodesic circular orbits which exist only for $r>r_{ph}^{(-)}$. To describe these figures it is useful to compare their curves with the curves in Figs. 3--5 for Schwarzschild's background from \cite{Pl3}. Thus, the lower curve in Fig. 6 for $a=0.015M$ is close to the corresponding Schwarzschild's curve,
whereas the upper curve in Fig. 6 only appears  due to the nonzero value $a$. Per the solutions of Eqs.  (\ref{16}) and (\ref{18}), the necessary value $\gamma$ tends to $\infty$ if $a\to 0$, similarly to the case of the upper curve in Fig. 1. If the value $a$ is greater than $0.015M$, both the curves in Fig. 7 differ from the corresponding curves in Fig. 6. Figures 8 and 9 describe the  situations for  $a=0.7M$: the first of them presents the dependence of the Lorentz factor on $r$ in the narrow space domain near $r_{ph}^{(-)}$, and the second is valid for values of $r$ less than those in Fig. 8.
\begin{figure}
[h] \centering
\includegraphics[width=5cm]{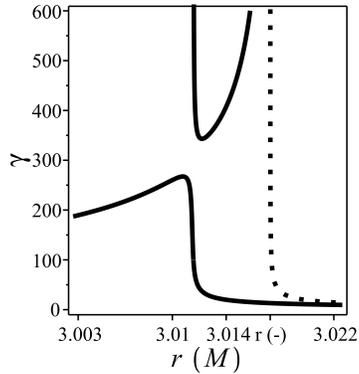}
\caption{\label{6}  Dependence of the Lorentz factor on $r$ for the highly relativistic circular orbits with $d\varphi/ds<0$ of the spinning particle in Kerr's background at $a=0.015M$, $\varepsilon_0=0.01$ (solid lines). }
\end{figure}

\begin{figure}
[h] \centering
\includegraphics[width=5cm]{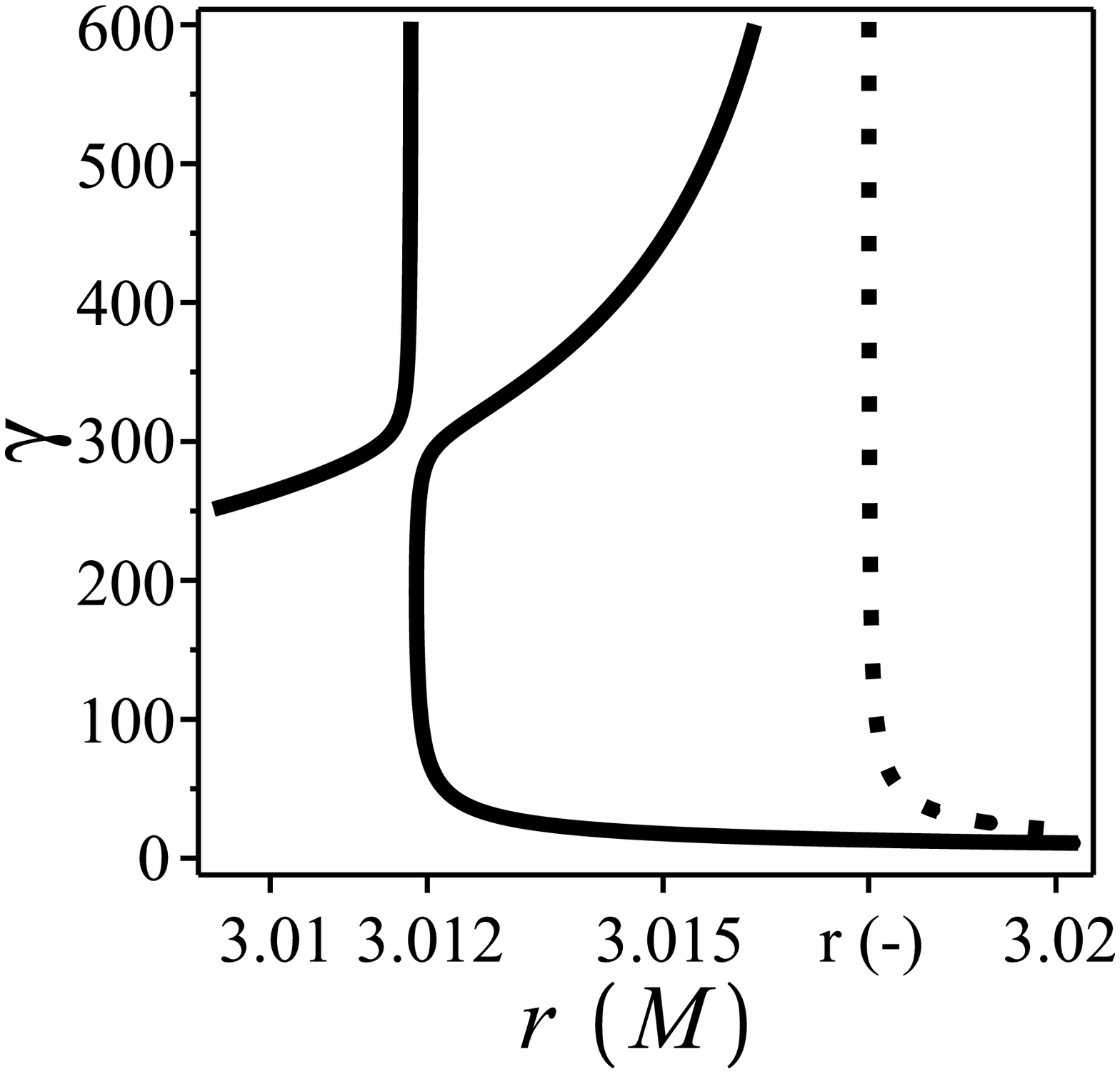}
\caption{\label{7}  Dependence of the Lorentz factor on $r$ for the highly relativistic circular orbits with $d\varphi/ds<0$ of the spinning particle in Kerr's background at $a=0.0153M$, $\varepsilon_0=0.01$. }
\end{figure}

\begin{figure}
[h] \centering
\includegraphics[width=5cm]{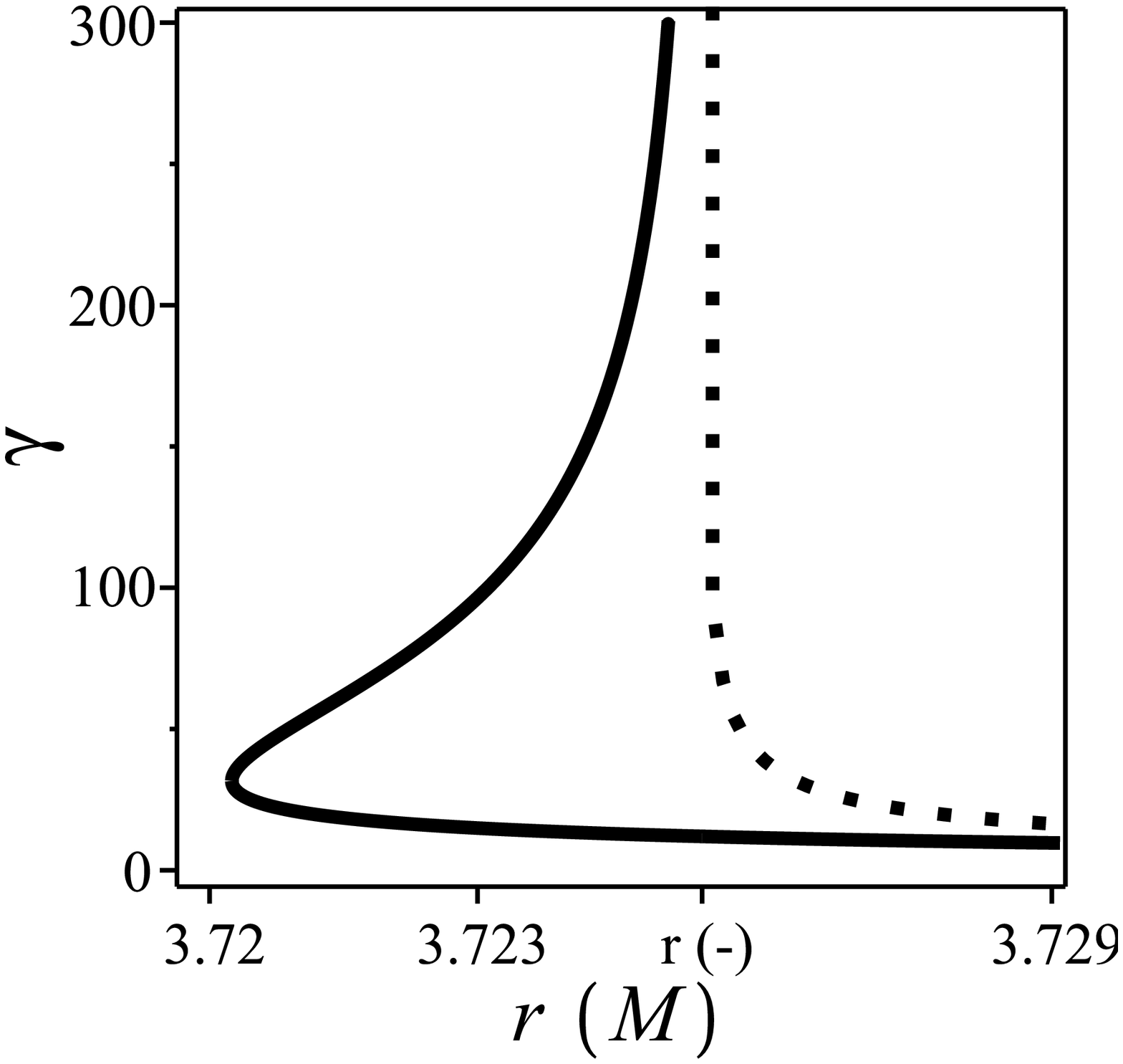}
\caption{\label{8} Lorentz factor  for the highly relativistic circular orbits with $d\varphi/ds<0$ of the spinning particle in Kerr's background at $a=0.7M$, $\varepsilon_0=0.01$ for $r$ close to $r_{ph}^{(-)}$. }
\end{figure}

\begin{figure}
[h] \centering
\includegraphics[width=5cm]{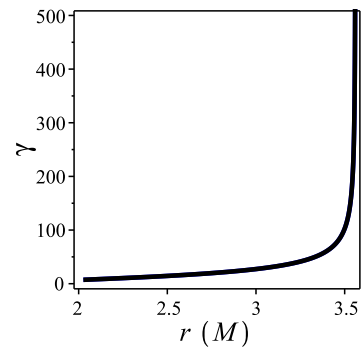}
\caption{\label{9}   Lorentz factor  for the highly relativistic circular orbits with $d\varphi/ds<0$ of the spinning particle in Kerr's background at $a=0.7M$, $\varepsilon_0=0.01$ for $r$ less than in Fig. 8.}
\end{figure}

\section{ Circular orbits with $a>0$, $S_\theta<0$ }

In this section we continue the presentation of some typical highly relativistic circular orbits of a spinning particle in the Kerr background which follow from Eqs. (16) and (18).
In contrast to Figs. 1--9,  Figs. 10--14 all describe different cases with the negative value $\varepsilon_0=-0.01$.

\subsection{Case with $d\varphi/ds>0$ (co-rotation)}

Figures 10 and 11 correspond to the same value $a=0.1M$ with the difference that Fig. 10 shows the dependence $\gamma$ on $r$ in the narrow space region near $r_{ph}^{(+)}$ whereas the graph in Fig. 11 is valid for $r$ beyond this region, when $r<r_{ph}^{(+)}$. Figure 12 at $a=0.5M$ is similar to Fig. 10.

\begin{figure}
[h] \centering
\includegraphics[width=5cm]{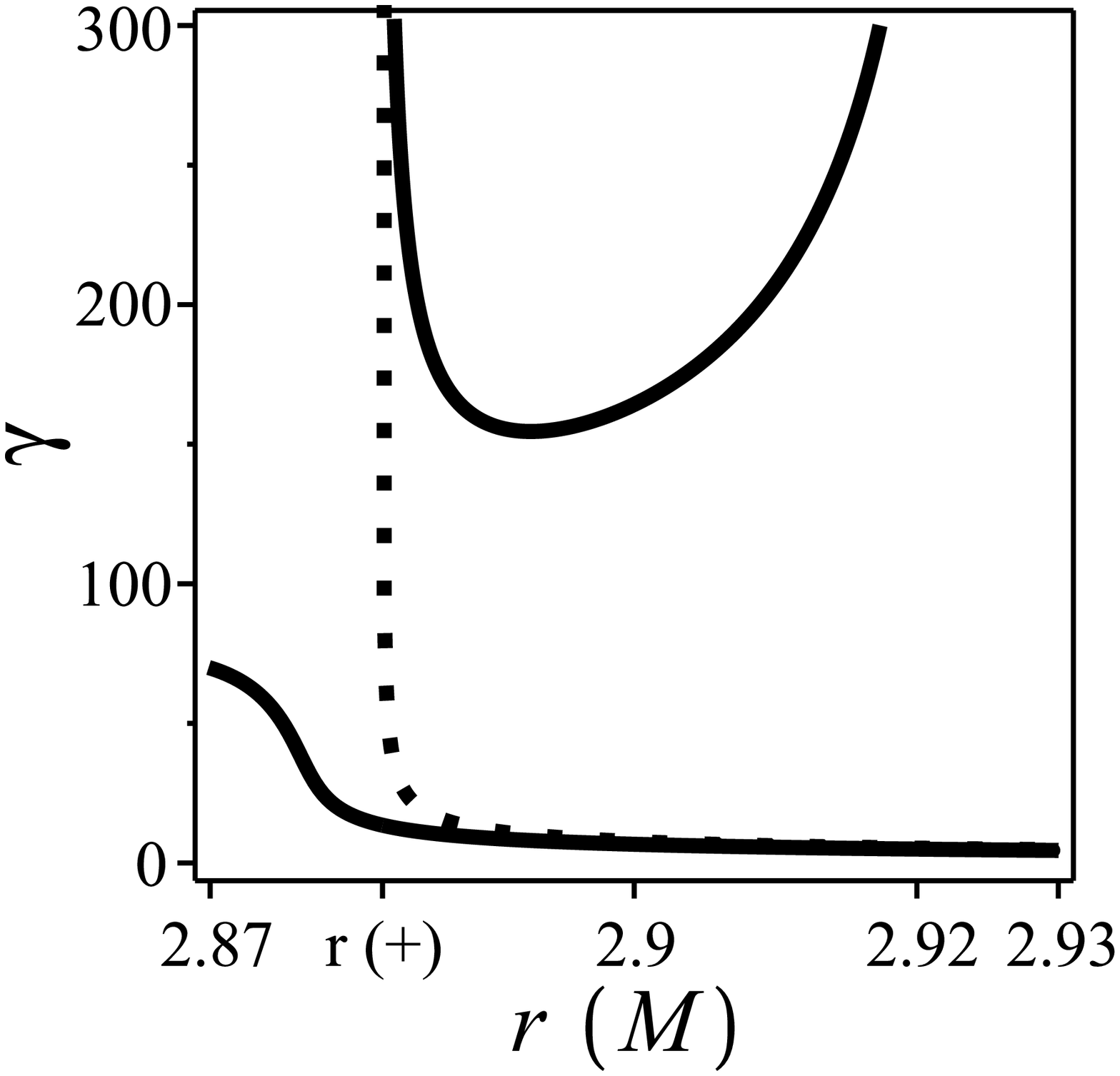}
\caption{\label{10}  Lorentz factor  for the highly relativistic circular orbits with $d\varphi/ds>0$ of the spinning particle in Kerr's background at $a=0.1M$, $\varepsilon_0=-0.01$ for $r$ close to $r_{ph}^{(+)}$. }
\end{figure}

\begin{figure}
[h] \centering
\includegraphics[width=5cm]{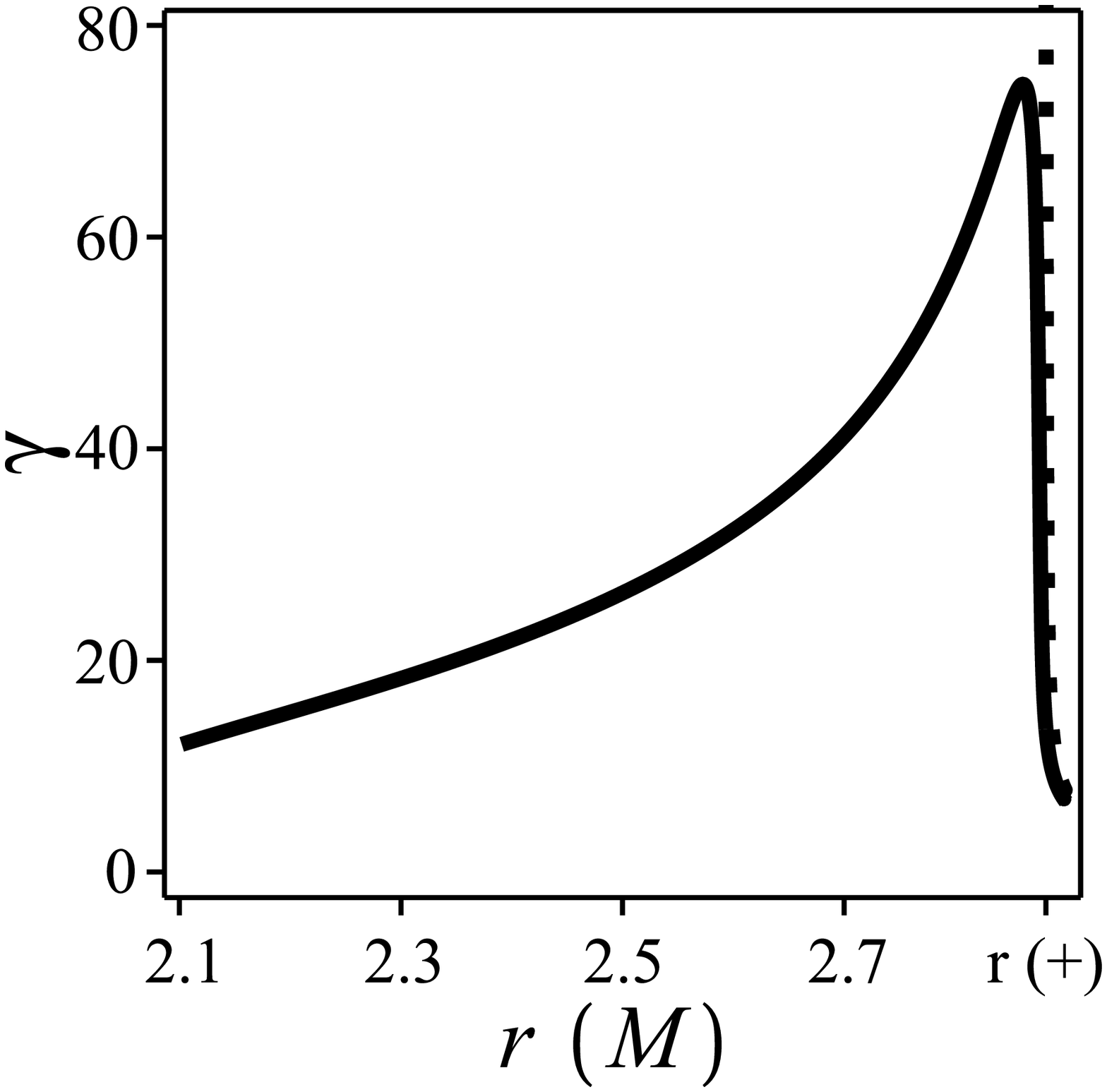}
\caption{\label{11}  Lorentz factor  for the highly relativistic circular orbits with $d\varphi/ds>0$ of the spinning particle in Kerr's background at $a=0.1M$, $\varepsilon_0=-0.01$ for $r$ less than in Fig. 10. }
\end{figure}

\begin{figure}
[h] \centering
\includegraphics[width=5cm]{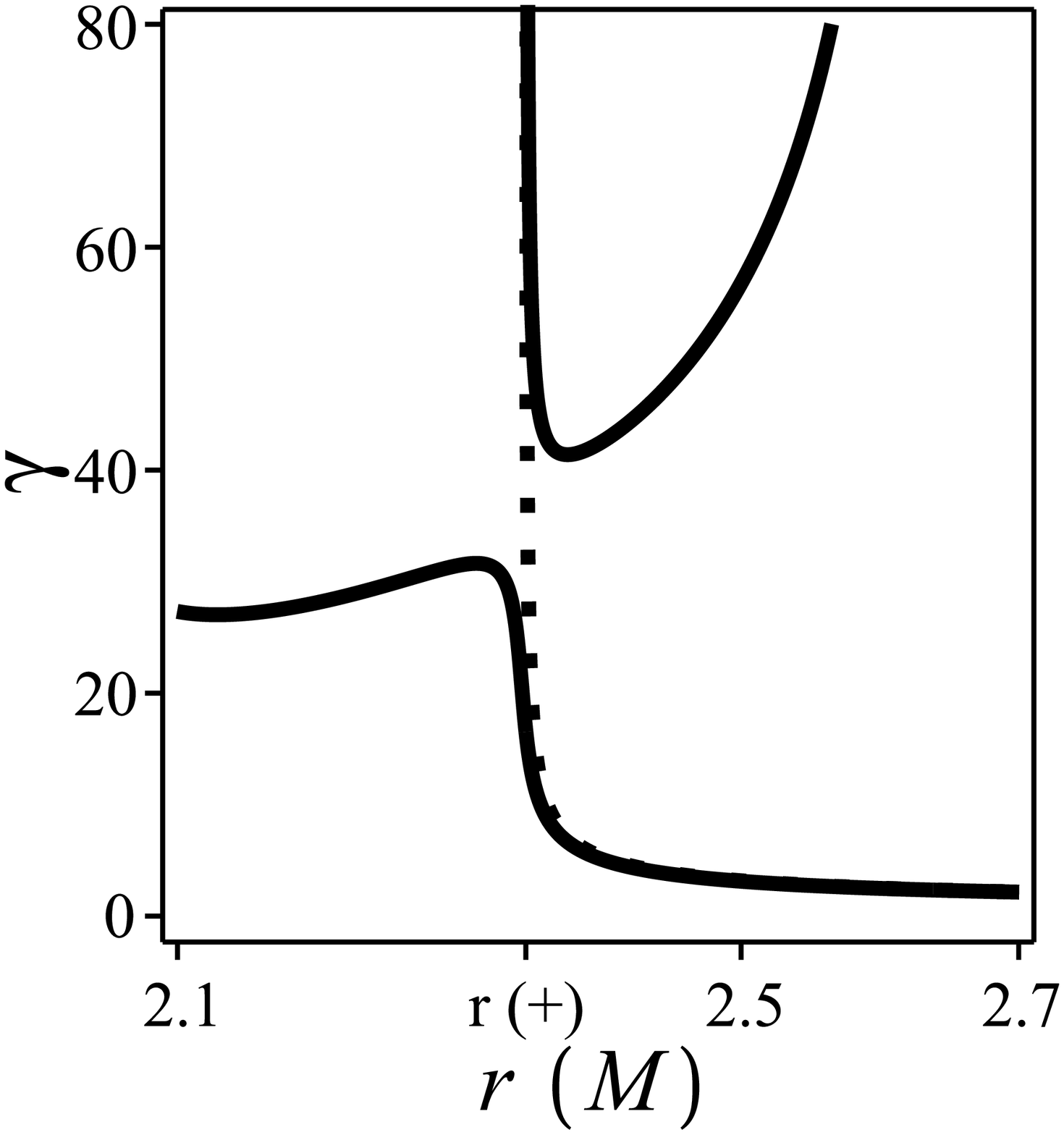}
\caption{\label{12}  Dependence of the Lorentz factor on $r$ for the highly relativistic circular orbits with $d\varphi/ds>0$ of the spinning particle in Kerr's background at $a=0.5M$, $\varepsilon_0=-0.01$. }
\end{figure}

\subsection{Case with $d\varphi/ds<0$ (counter-rotation)}

Figures 13 and 14 describe the possible highly relativistic circular orbits which exist both for $r>r_{ph}^{(-)}$ and $r<r_{ph}^{(-)}$ for different values $a$:  $0.15M$ and $M$, respectively. Note the left-hand curves in these figures are caused  by the nonzero Kerr parameter $a$ only; i.e., similar orbits are absent in Schwarzschild's field.

\begin{figure}
[h] \centering
\includegraphics[width=5cm]{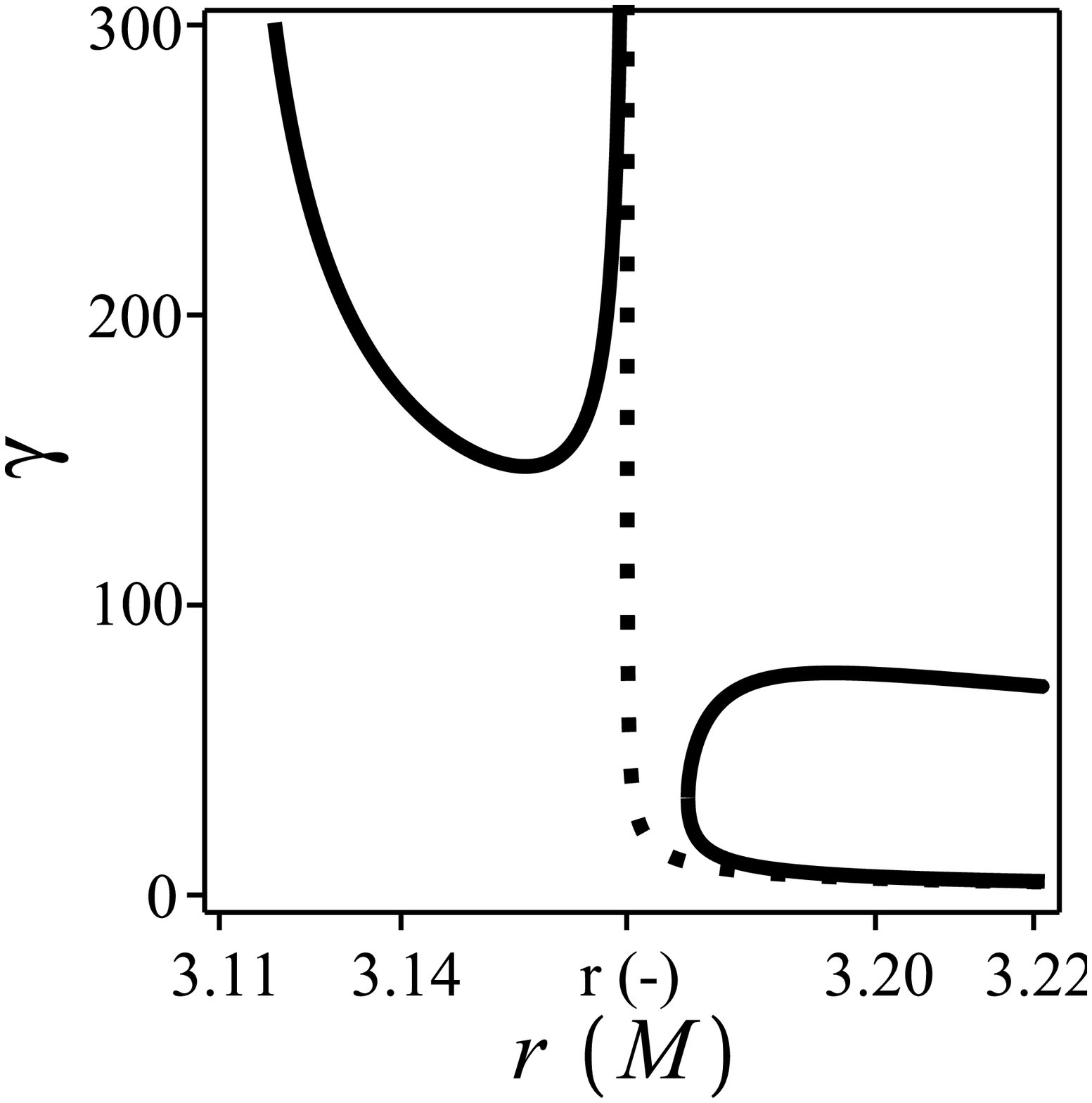}
\caption{\label{13}  Dependence of the Lorentz factor on $r$ for the highly relativistic circular orbits with $d\varphi/ds<0$ of the spinning particle in Kerr's background at $a=0.15M$, $\varepsilon_0=-0.01$. }
\end{figure}

\begin{figure}
[h] \centering
\includegraphics[width=5cm]{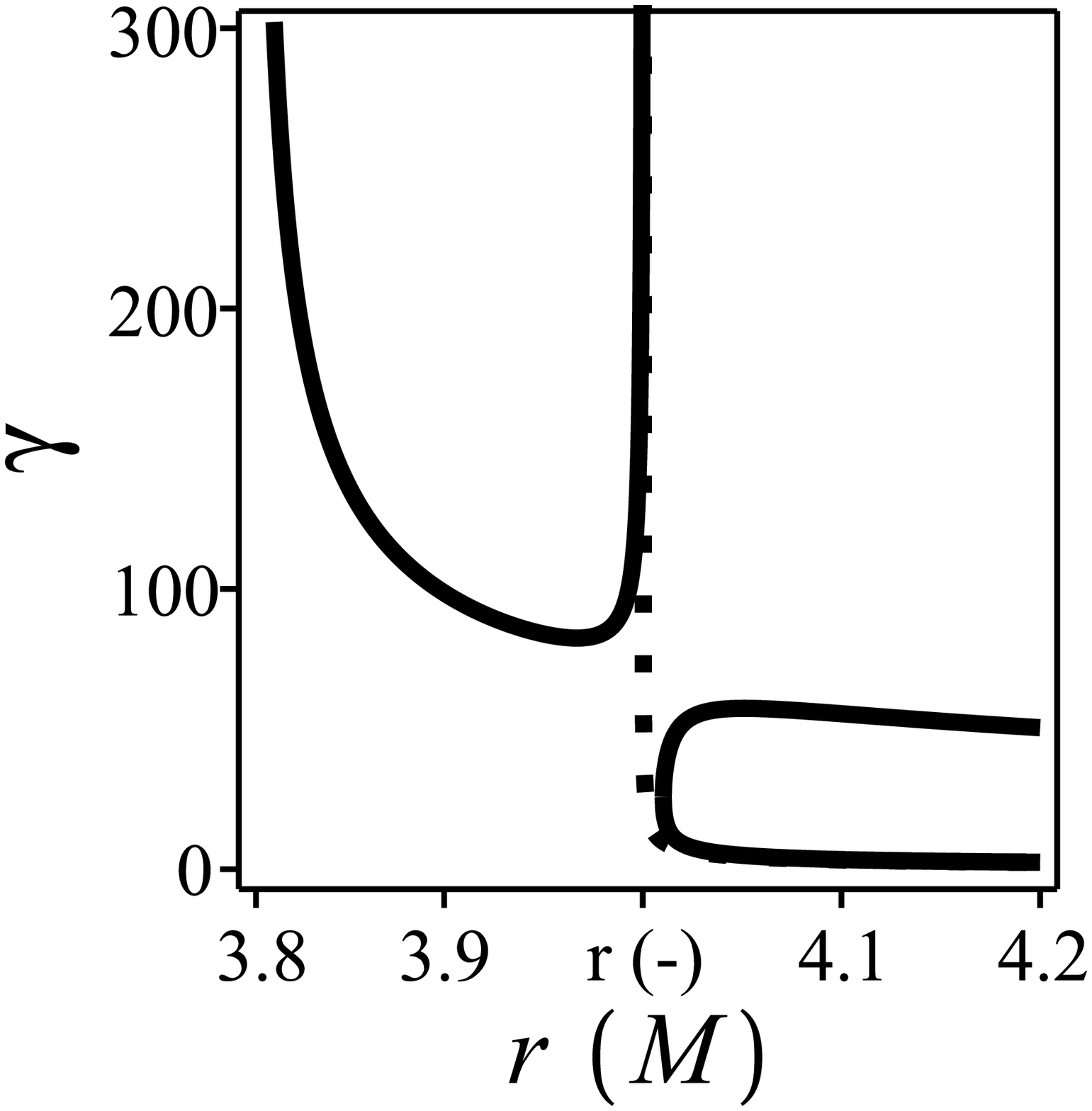}
\caption{\label{14} Dependence of the Lorentz factor on $r$ for the highly relativistic circular orbits with $d\varphi/ds<0$ of the spinning particle in Kerr's background at $a=M$, $\varepsilon_0=-0.01$.  }
\end{figure}

\begin{figure}
[h] \centering
\includegraphics[width=5cm]{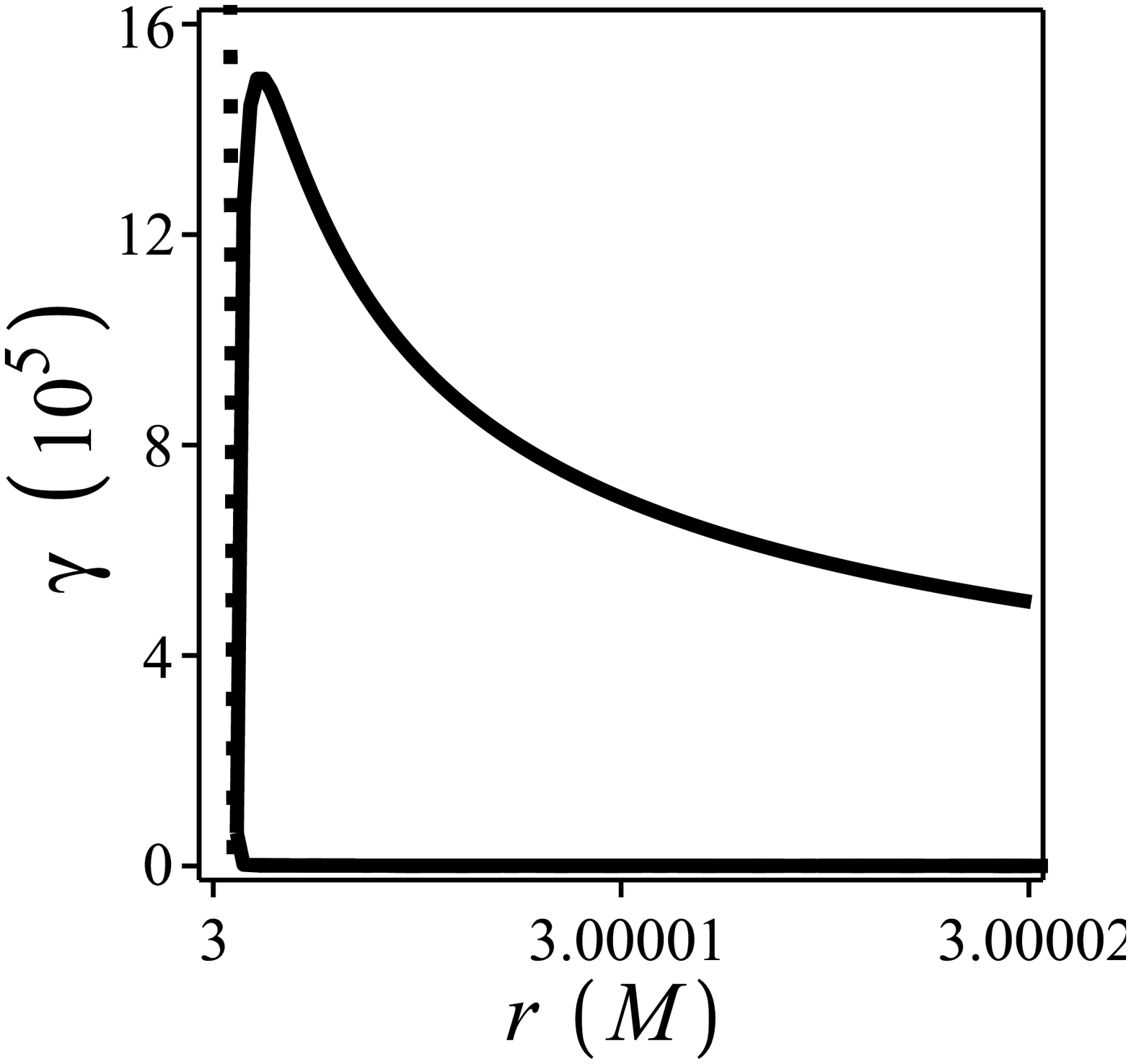}
\caption{\label{15} Lorentz factor vs $r$ for $\varepsilon_0=10^{-6}$, $d\varphi/ds>0$, $a=0$. }
\end{figure}

\begin{figure}
[h] \centering
\includegraphics[width=5cm]{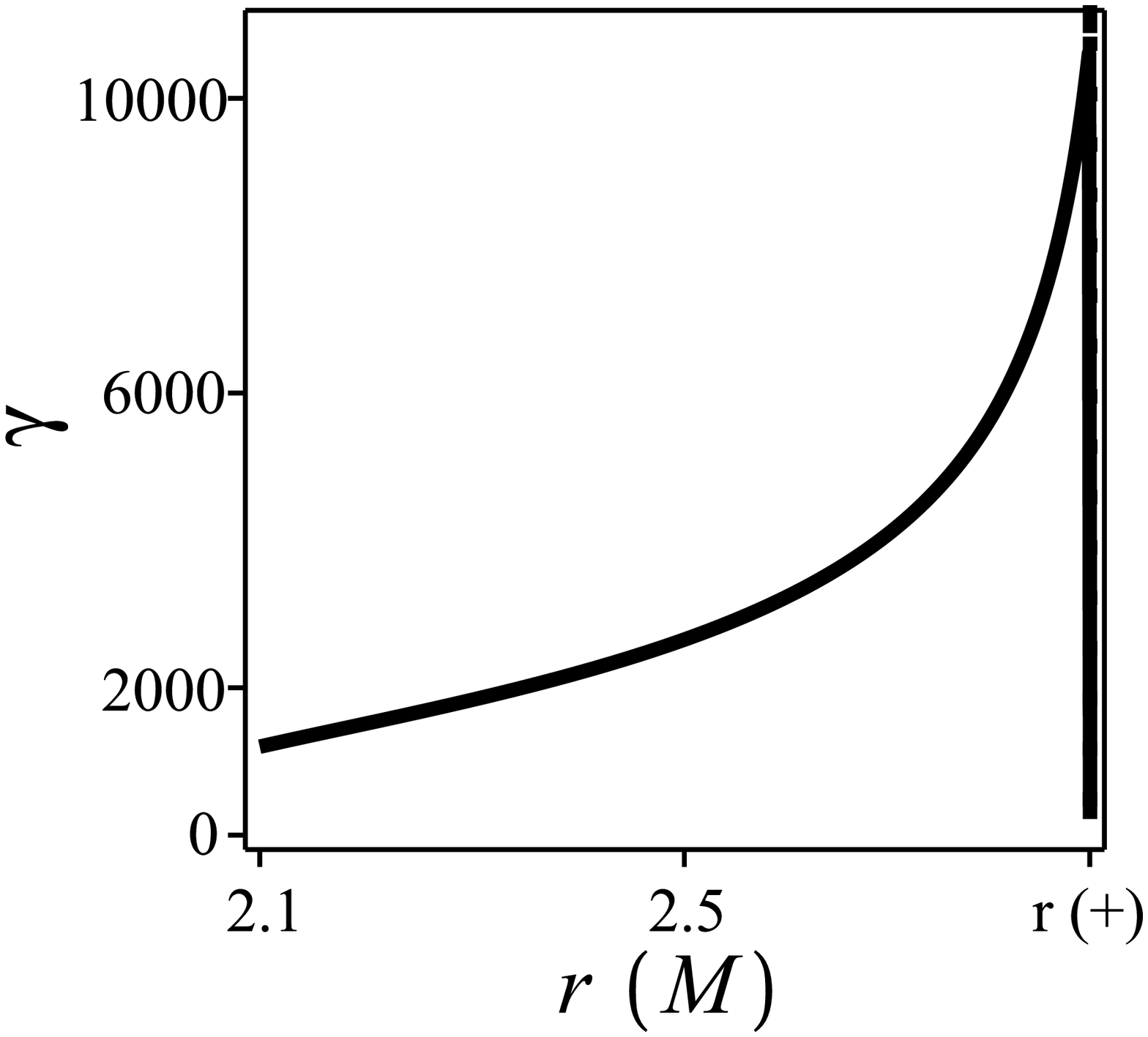}
\caption{\label{16}  Lorentz factor vs $r$ for $\varepsilon_0=-10^{-6}$, $d\varphi/ds>0$, $a=0.1M$. }
\end{figure}

\section{Conclusions }

Let us summarize the data which we obtained from Eqs. (\ref{16}) and (\ref{18}) concerning the possible highly relativistic circular orbits of a spinning particle as described by the MP Eqs. (\ref{1}) and (\ref{2}) in Kerr's background. First, we begin with the orbits in the space region far from $r_{ph}^{(-)}$. These orbits exist if the signs of $\varepsilon_0$ and
$d\varphi/ds$ are the same, both positive or negative for the chosen positive $a$. According to our Fig. 5  and per Fig. 2 from \cite{Pl3}, the values of the particle's $\gamma$-factor which are necessary for the realization of these orbits at different $a$ for the large $r$ are practically the same as for Schwarzschild's background. That is, in this case the role of the parameter $a$ is negligible. Note that, for any fixed large $r$ (as compare to $r_{ph}^{(-)}$), there are two possibilities for the spinning particle motion on the circular orbits with this $r$, (i) If $\gamma$ is close to 1, then the motion practically coincides with the geodesic nonhighly relativistic motion; i.e., the spin-gravity effects are very small, (ii) If $\gamma\gg 1$,  due to the essential spin-gravity effect only, the spinning particle feels the great additional attractive action which preserves the fast motion of this particle in some direction with growing $r$, as it takes place for the fast spinless particle.

Figures 1--3 and 6--9, and others show the situations with the circular orbits in the space domain between $r_{ph}^{(+)}$ and $r_{ph}^{(-)}$ when the parameter $a$ plays an important role. Figures 10--12 describe the orbits with $r<r_{ph}^{(-)}$ which in certain sense is similar to the circular orbits in the region $2m<r<3M$ of Schwarzschild's background (see Figs. 4 and 5 in \cite{Pl3}). In contrast to the orbits in the region $r>r_{ph}^{(-)}$ which are possible due to the attractive action of the spin-gravity coupling on the particle, in the region $r<r_{ph}^{(-)}$ there are  circular orbits which are caused by both the significant attractive and repulsive actions.

All Figs. 1--14 correspond to the $|\varepsilon_0|=10^{-2}$. Naturally, for other values of $|\varepsilon_0|$ the form of the corresponding graphs is changed, but many features are similar. Mainly, $\gamma$-factor is proportional to $1/\sqrt{|\varepsilon_0|}$; i.e., it is growing with decreasing
$|\varepsilon_0|$. Figures 15 and 16 give the two concrete examples.

Certainly, the pictures when a spinning particle or a spinless particle remains indefinitely on the highly relativistic circular orbits in Kerr's background hold in the ideal case, when perturbations are neglected, because these orbits are unstable. In reality, one can deal with fragments of trajectories close to the corresponding circular orbits.

Do some particles in cosmic rays posses the sufficiently high $\gamma$-factor for motions on the highly relativistic circular orbits, or on some their fragments, in the gravitational field of a Kerr black hole, which are considered above? Yes, they do.
Per the numerical estimates similar to those from \cite{Pl2}, for an electron in the gravitational field of a black hole with three times the Sun's mass, the value $|\varepsilon_0|$ is equal $4\times 10^{-17}$. Then the necessary value of the $\gamma$-factor for the realization of some highly relativistic circular orbits by the electron near this black hole is of the order $10^8$. This $\gamma$-factor corresponds to the energy of the electron free motion of the order $10^{14}$ eV. Analogously, for a proton in the field of such a black hole, the corresponding energy is of the order  $10^{18}$ eV. For a massive black hole those values are greater: for example, if $M$ is equal to $10^6$ times the Sun's mass, the corresponding value of the energy for an electron is of the order $10^{17}$ eV and for a proton it is $10^{21}$ eV.
Naturally, far from the black hole, if $r\gg r_{ph}^{(-)}$, these values are greater because the necessary $\gamma$-factor is proportional to $\sqrt{r}$.

Note that for a neutrino near the black hole with three times the Sun's mass, the necessary values of its $\gamma$-factor for motions on the highly relativistic circular orbits correspond to the neutrino's energy of the free motion of the order of $10^5$ eV. If the black hole's mass is of the order of $10^6$ of the Sun's mass, the corresponding value is of the order of $10^8$ eV.

Can the highly relativistic spin-gravity effects be registered by the observation of the electromagnetic synchrotron radiation from some black holes? It may be so; however, it is difficult to determine the situation when the circular orbits of a spinning charged particle and its synchrotron radiation are caused by the magnetic field or when they are caused by the spin-gravity coupling. A detailed analysis of the observational data is necessary.

\end{document}